\documentclass[aps,pra,twocolumn,showpacs,superscriptaddress]{revtex4-1}

\usepackage{amsfonts,mathrsfs,amsmath,amsthm,amssymb}
\usepackage{bm,times}
\usepackage{graphicx,epsfig}
\usepackage[colorlinks=true,breaklinks=true,linkcolor=blue,citecolor=blue]{hyperref}

\def \ket {\rangle}
\def \bra {\langle}

\def \non {\nonumber}
\def \im {\rm i}

\newcommand{\be}{\begin{eqnarray}}
\newcommand{\ee}{\end{eqnarray}}

\makeatletter
    
    \newcommand{\Rmnum}[1]{\expandafter\@slowromancap\romannumeral #1@}
\makeatother

\hypersetup{
  pdfauthor={Tao Zhou},
  pdfkeywords={},
  pdftitle={Success probabilities for universal unambiguous discriminators between unknown pure states},
  pdfsubject = {Research Paper},
  pdfpagemode = UseNone
}

\begin{document}

\title{Success probabilities for universal unambiguous discriminators between unknown pure states}

\author{Tao Zhou}
\email{taozhousc@gmail.com}
\affiliation{Quantum Optoelectronics Laboratory, School of Physics and Technology, Southwest Jiaotong University, Chengdu 610031, China}

\date{\today}

\begin{abstract}
A universal programmable discriminator can perform the discrimination between two unknown states, and the optimal solution can be approached via the discrimination between the two averages over the uniformly distributed unknown input pure states, which has been widely discussed in previous works. In this paper, we consider the success probabilities of the optimal universal programmable unambiguous discriminators when applied to the pure input states. More precisely, the analytic results of the success probabilities are derived with the expressions of the optimal measurement operators for the universal discriminators and we find that the success probabilities have nothing to do with the dimension $d$ while the amounts of the copies in the two program registers are equal. The success probability of programmable unambiguous discriminator can asymptotically approach to that of usual unambiguous discrimination (state comparison) as the number of copies in program registers (data register) goes to infinity.
\end{abstract}

\maketitle

The discrimination between quantum states is a basic tool in quantum information processing, and this is a nontrivial problem since an unknown state cannot be perfectly cloned~\cite{non-cloning}. Usually, there are two basic strategies to achieve the state discrimination: Minimum-error discrimination (MD)~\cite{Helstrom,Holevo,ME} with a minimal probability for the error, and unambiguous discrimination (UD)~\cite{UD} with a minimum probability of inconclusive results. In those works, a quantum state is chosen from a set of known states but we do not know which and want to determine the actual states. 

The discrimination problems above are dependent on the set of states to be distinguished, and the device for the discrimination is not universal but specifically designed for the states. As in the spirit of programmable quantum devices~\cite{programmable}, it is interesting to design a discrimination device that does not need to change as the input states change. Such a universal device that can unambiguously discriminate between two unknown qubit states has first been constructed by Bergou and Hillery~\cite{PRL94.160501}. In this programmable quantum device, two possible states enter two program registers as ``programs" respectively, and the data register is prepared with a third state (guaranteed to be one of the two possible states) which one wishes to identify. One amazing feature of this discriminator is that the states in the device can be unknown, which means no classical knowledge on the states is provided, and it is capable to distinguish any pair of states in this device.

Later, the generalizations and the experimental realizations of programmable discriminators are introduced and widely discussed~\cite{pure,QIC12.1017,PRA73.012328,PRA82.042312,generalization,PRA88.052304}. The problems with multiple copies in program and data registers or with high-dimensional states in the registers are considered. Furthermore, the case that each copy in the registers is the mixed state is also treated~\cite{PRA82.042312}. In most of these literatures, either the unambiguous discrimination or the minimum-error scheme is used,  and quite recently, the two strategies are unified in Ref.~\cite{PRA88.052304} by introducing an error margin. With multi-copy and high-dimensional pure states in the registers, the optimal solution and success probability are hard to obtain. However, by taking the average of the unknown states, the task is equivalent to the discrimination of known mixed states in previous works. Because the equivalent mixed states are highly symmetrical objects, the success probabilities for discrimination between them, or the average success probabilities (ASPs) can be derived and then the optimal solution is obtained using the symmetry properties. The ASPs can be used to determining the optimality of the discriminators, or namely, the discriminators with the maximal ASPs are optimal. In fact, as mentioned before, the unknown pure states instead of the average mixed states are distinguished in most cases, and the ASP is not the success probability when the device works. Therefore, it is useful and meaningful to obtain the success probabilities for the programmable discrimination between the pure states, or pure success probabilities (PSPs) for short. Moreover, if we have the PSP, take the average of it and we can obtain the ASP. Unfortunately, except the results for few special simple cases~\cite{PRL94.160501,pure}, PSPs for the general case with multi-copy and high-dimensional states are left behind and still not obtained.

The main purpose of this paper is to evaluate the PSPs of the optimal universal unambiguous discriminators between two unknown pure states for the general case. Following the optimal measurement operators for the universal unambiguous discriminator in Ref.~\cite{QIC12.1017}, we show that the PSPs are related to Wigner's $D$ function and the Clebsch-Gordan (CG) coefficients for the angular momentum coupling. The analytic results of PSPs can be obtained with the exact expressions of Wigner $D$ function and the CG coefficients.

\textit{Universal programmable unambiguous discriminators.---}
The general case of the programmable discriminators has been systematically studied~\cite{QIC12.1017}, and the optimal solutions are obtained with the representation theory of U($d$) group and Jordan basis method. The discriminator consists of two program registers $A$ and $C$, and one data register $B$. It is assumed that systems $A$ and $C$ are prepared in the states $|\phi_1\ket$ and $|\phi_2\ket$, each with $n_A$ and $n_C$ copies respectively, and system $B$ is prepared in $n_B$ copies of either $|\phi_1\ket$ or $|\phi_2\ket$, with {\it a priori} probabilities $\eta_1$ and $\eta_2$, such that $\eta_1+\eta_2=1$. Such a device can measure and then may distinguish the total input states
\be
\label{states}
|\Phi_1\ket&=&|\phi_1\ket_A^{\otimes n_A}|\phi_1\ket_B^{\otimes n_B}|\phi_2\ket_C^{\otimes n_C},\non\\
|\Phi_2\ket&=&|\phi_1\ket_A^{\otimes n_A}|\phi_2\ket_B^{\otimes n_B}|\phi_2\ket_C^{\otimes n_C},
\ee
with the unknown qudit states $|\phi_1\ket$ and $|\phi_2\ket$ in $d$-dimensional Hilbert space $\mathcal{H}$. Mathematically, this programmable discriminator is defined by the elements of a universal POVM $\{\Pi_1,\Pi_2,\Pi_0\}$, where $\Pi_{1}(\Pi_2)$ is associated with the input state $|\Phi_1\ket(|\Phi_2\ket)$, and $\Pi_0$ corresponds to an inconclusive result.  Without loss of generality, we assume that $n_A\geqslant n_C$, and set $n_1=n_A+n_B,n_2=n_B+n_C,N=n_A+n_B+n_C$.

It is not difficult to see that the state $|\Phi_1\ket$ lies in the tensor space $\mathbb{H}_1=\mathcal{H}^{[n_1]}\otimes\mathcal{H}^{[n_c]}$, while $|\Phi_2\ket$ in $\mathbb{H}_2=\mathcal{H}^{[n_A]}\otimes\mathcal{H}^{[n_2]}$~\cite{explain}. The irreducible  bases of $\mathcal{H}^{[n_1]}$ and $\mathcal{H}^{[n_c]}$ can be coupled to give another irreducible basis $\bigg|\begin{array}{c} [\nu_1]\\  \omega_1\end{array}\bigg\ket_1$  by the representation theory of U($d$) group. $\bigg|\begin{array}{c} [\nu_1]\\  \omega_1\end{array}\bigg\ket_1$ form the complete orthogonal basis of $\mathbb{H}_1$, where all the possible Young diagrams for $[\nu_1]$ constitute a set $S_1=\{[N-k,k]: k=0,1,...,n_C\}$ and $\omega_1=1,2,\cdots,d^{[\nu_1]}$ with $d^{[\nu_1]}$ the dimension of the irreducible space labeled by $[\nu_1]$. Similarly, one has the complete orthogonal basis $\bigg|\begin{array}{c} [\nu_2]\\  \omega_2\end{array}\bigg\ket_2$ for $\mathbb{H}_2$, and besides the Young diagrams in $S_1$, $[\nu_2]$ can take ones in the set $S_2=\{[N,N-k]: k=n_C+1,n_C+2,...,\min(n_A,n_2)\}$ when $n_A>n_B$. Furthermore, one has the decomposition $\mathbb{H}_2=H_2\oplus H_2^\bot$, where the subspace $H_2$ is spanned by all the bases $\bigg|\begin{array}{c} [\nu_2]\\  \omega_2\end{array}\bigg\ket_2$ with $[\nu_2]\in S_1$ while the subspace $H_2^\bot$ by the ones with $[\nu_2]\in S_2$. $H_2^\bot$ is orthogonal to both $\mathbb{H}_1$ and $H_2$, and the basis $\bigg\{\bigg|\begin{array}{c} [\nu_1]\\  \omega_1\end{array}\bigg\ket_1:[\nu_1]\in S_1,\omega_1=1,2,...,d^{[\nu_1]}\bigg\}$ for $\mathbb{H}_1$ and the basis $\bigg\{\bigg|\begin{array}{c} [\nu_2]\\  \omega_2\end{array}\bigg\ket_2:[\nu_2]\in S_1,\omega_2=1,2,...,d^{[\nu_2]}\bigg\}$ for $H_2$ form the Jordan basis~\cite{QIC12.1017,PRA73.032107}.

Now, we can introduce the optimal measurement operators for the universal programmable unambiguous discriminators as follows
\be
\label{operators}
\Pi_1&=&\sum_{k=1}^{n_C}\sum_{\omega=1}^{d_k} \frac{1-q_{k,1}^{\rm opt}}{1-O_k^2}|\psi_{k,\omega}^\bot\ket_2\bra\psi_{k,\omega}^\bot|,\non\\
\Pi_2&=&\sum_{k=1}^{n_C}\sum_{\omega=1}^{d_k}\frac{1-q_{k,2}^{\rm opt}}{1-O_k^2}|\psi_{k,\omega}^\bot\ket_1\bra\psi_{k,\omega}^\bot|+\openone^\bot,\non\\
\Pi_0&=&\openone-\Pi_1-\Pi_2,
\ee
where $d_k=d^{[N-k,k]}$, $\openone^\bot$ and $\openone$ are the identity operators on $H_2^\bot$ and  $\mathbb{H}=\mathbb{H}_1\cup\mathbb{H}_2$, respectively. $|\psi_{k,\omega}^\bot\ket_1$ and $|\psi_{k,\omega}^\bot\ket_2$ are normalized orthogonal vectors to $\bigg|\begin{array}{c} [\nu]\\  \omega\end{array}\bigg\ket_1$ and $ \bigg|\begin{array}{c} [\nu]\\  \omega\end{array}\bigg\ket_2$ respectively in the subspace spanned by them, and $O_k$ are the inner products of Jordan basis dependent only on the Young diagram $[\lambda]=[N,N-k]$. The parameter $q_{k,1}^{\rm opt}$ is taken as
\be
\label{parameter}
q_{k,1}^{\rm opt}=\left\{\begin{array}{ll} 1 & \textrm{for } \eta_1<c_k\\
\sqrt{\frac{\eta_2d_1}{\eta_1d_2}}O_k & \textrm{for } c_k\leqslant\eta_1\leqslant d_k\\
O_k^2 & \textrm{for } \eta_1> d_k\end{array}\right.,
\ee
$q_{k,2}^{\rm opt}=O_k^2/q_{k,1}^{\rm opt}$ and the boundaries $c_k=d_1O_k^2/(d_2+d_1O_k^2)$ and $d_k=d_1/(d_1+d_2O_k^2)$.
Details on the derivation of the optimal measurement operators for the universal programmable discriminator are discussed in Ref.~\cite{QIC12.1017}.  Now we can come to the PSPs in the next section.

\textit{PSPs for universal unambiguous discriminators.---}
Since the unambiguous discriminators are universal, the optimal operators $\Pi_0$, $\Pi_1$, and $\Pi_2$  are applicable for the PSPs to the discrimination between the pure states $|\Phi_1\ket$ and $|\Phi_2\ket$. Armed with the expressions of the optimal operators in Eq.~(\ref{operators}), the optimal success probability reads
\be
\label{SuccP}
P&=&\eta_1\bra\Phi_1|\Pi_1|\Phi_1\ket+\eta_2\bra\Phi_2|\Pi_2|\Phi_2\ket
\ee
To give the exact expression of PSP, let us consider $\bra\Phi_1|\Pi_1|\Phi_1\ket$ first. With the expression of $\Pi_1$ and the relationship
\be
\bra\Phi_1|\psi_{k,\omega}^\bot\ket_2\bra\psi_{k,\omega}^\bot|\Phi_1\ket=(1-O_k^2)\bigg\bra\Phi_1\bigg|\begin{array}{c} [\lambda]\\  \omega\end{array}\bigg\ket_1\bigg\bra\begin{array}{c} [\lambda]\\  \omega\end{array}\bigg|\Phi_1\bigg\ket\non,
\ee
we have
\be
\label{expectaion1}
\bra\Phi_1|\Pi_1|\Phi_1\ket=\sum_{[\lambda]\in S_1}(1-q_{k,1}^{\rm opt})\bra\Phi_1|\openone^{[\lambda]}|\Phi_1\ket,
\ee
where $\openone^{[\lambda]}=\sum_{\omega=1}^{d^{[\lambda]}}\bigg|\begin{array}{c} [\lambda]\\  \omega\end{array}\bigg\ket_1\bigg\bra\begin{array}{c} [\lambda]\\  \omega\end{array}\bigg|$
is the identity operator on the irreducible representation space labeled by $[\lambda]=[N,N-k]$. The following lemma is very useful to calculate the PSPs.

\textit{Lemma.}
The expectation of the operator $\openone^{[\lambda]}$ with respect to $|\Phi_1\ket$ can be expressed as
\be
\label{CG}
\bra\Phi_1|\openone^{[\lambda]}|\Phi_1\ket=\sum_M\bigg|\sum_lD^{(\frac{n_C}{2})}_{l\frac{n_C}{2}}C^{\frac{N}{2}-k,M}_{\frac{n_1}{2}\frac{n_1}{2},\frac{n_C}{2}l}\bigg|^2,
\ee 
where $D^{(\frac{n_C}{2})}_{l\frac{n_C}{2}}$ are the Wigner $D$ function~\cite{Edmonds}, and $C^{\frac{N}{2}-k,M}_{\frac{n_1}{2}\frac{n_1}{2},\frac{n_C}{2}l}$ are the CG coefficients~\cite{explain2}.

For readability, we postpone the detailed proof of this lemma to the technical appendix. Plugging the exact expressions of CG coefficients and Wigner $D$ function into Eq.~(\ref{CG}) and by some algebra, we have
\be
\bra\Phi_1|\openone^{[\lambda]}|\Phi_1\ket&=&\frac{(N-2k+1)n_1!n_C!}{k!(N-k+1)!}(\sin^2\frac{\beta}{2})^{n_C}\non\\
&\times&{}_2F_1(n_1-k+1,k-n_C;1;-\cot^2\frac{\beta}{2}),\non
\ee
where ${}_2F_1(a,b;c;z)$ is the Gauss hypergeometric function~\cite{Wongbook}. Finally, we can obtain
\be
\eta_1\bra\Phi_1|\Pi_1|\Phi_1\ket=a\big(1-|\bra\phi_1|\phi_2\ket|^2\big)^{n_C},\non
\ee
where the relationship $\sin^2\frac{\beta}{2}=1-|\bra\phi_1|\phi_2\ket|^2$ has been used and
\be
a&=&\eta_1\sum_{k=1}^{n_C}(1-q_{k,1}^{\rm opt})\frac{(N-2k+1)n_1!n_C!}{k!(N-k+1)!}\non\\
&&\times{}_2F_1(n_1-k+1,k-n_C;1;-\cot^2\frac{\beta}{2}).\non
\ee
Similar discussions can be carried on for  $\bra\Phi_2|\Pi_2|\Phi_2\ket$, and one can also obtain
\be
\eta_2\bra\Phi_2|\Pi_2|\Phi_2\ket=(b+c)\big(1-|\bra\phi_1|\phi_2\ket|^2\big)^{n_2},\non
\ee
with
\be
b&=&\eta_2\sum_{k=1}^{n_C}(1-q_{k,2}^{\rm opt})\frac{(N-2k+1)n_2!n_A!}{k!(N-k+1)!}\non\\
&&\times{}_2F_1(n_A-k+1,k-n_2;1;-\cot^2\frac{\beta}{2}),\non\\
c&=&\eta_2\sum_{k=n_C+1}^{\min(n_A,n_2)}\frac{(N-2k+1)n_2!n_A!}{k!(N-k+1)!}\non\\
&&\times{}_2F_1(n_A-k+1,k-n_2;1;-\cot^2\frac{\beta}{2}).\non
\ee

Now, the exact values of PSPs can be expressed as
\be
\label{PSP}
P=a\big(1-|\bra\phi_1|\phi_2\ket|^2\big)^{n_C}+(b+c)\big(1-|\bra\phi_1|\phi_2\ket|^2\big)^{n_2}.
\ee
This is the main result in this paper, and in the following we will use it to discuss some specific cases.

\textit{Remarks and discussion.---}(i) From Eq.~(\ref{parameter}), it is easy to notice that the parameters $q_{k,1}^{\rm opt}$ and $q_{k,2}^{\rm opt}$ may be dependent on the dimension $d$ of the Hilbert space $\mathcal{H}$ unless $n_A=n_C$, and thus the values of PSPs are generally dependent on $d$. As the reference states $|\phi_1\ket$ and $|\phi_2\ket$ are completely unknown, there is no priority for one data register to own more copies than the other one, so we put $n_A=n_C=m,n_B=n$ in this section. The PSPs are now independent on the dimension $d$, which is very different from the cases for the averages of the input states, where the ASPs are always dependent on $d$. This is reasonable since the average of the overlap $|\bra\phi_1|\phi_2\ket|$ over the unknown states $|\phi_1\ket$ and $|\phi_2\ket$ is always related to $d$.

(ii) When the number of copies in data register goes to infinity while these in program registers are finite, from Eq.~(\ref{PSP}), we have the asymptotic limit of PSP for each finite $m$
\be
P_m\equiv\lim\limits_{n \to \infty}{P}=\sum_{k=1}^m a_{mk}(\cos^2\frac{\beta}{2})^{m-k}(\sin^2\frac{\beta}{2})^k,
\ee
where the coefficients $a_{mk}$ in each term can be calculated and if we further define $a_{00}=1$ and $a_{m0}=1$, the values of $a_{mk}(m\geqslant k\geqslant 0)$ can be listed in the following
\begin{center}
\begin{tabular}{cp{7mm}ccccccccccc}
$m=0:$ & & & & & & & 1\\
\noalign{\smallskip\smallskip} $m=1:$ & & & & & & 1 & & 1\\
\noalign{\smallskip\smallskip} $m=2:$ & & & & & 1 & & 2 & & 1\\
\noalign{\smallskip\smallskip} $m=3:$ & & & & 1 & & 3 & & 3 & & 1\\
\noalign{\smallskip\smallskip} $m=4:$ & & & 1 & & 4 & & 6 & & 4 & & 1\\
  \vdots & &\vdots&\vdots&\vdots&\vdots&\vdots&\vdots&\vdots&\vdots&\vdots&\vdots&\vdots\\
\end{tabular}
\end{center}

In the triangular array above, each entry happens to be the sum of the two upper entries, and this is the celebrated \emph{Pascal's triangle}, a geometric representation of the binomial coefficients. Hence the elements in the array above are the binomial coefficients, say $a_{mk}=\left(\begin{array}{c} m\\
k \end{array}\right)$. As a result,
\be
\label{ninfty}
P_m&=&\sum_{k=0}^m\left(\begin{array}{c} m\\
k \end{array}\right)(\cos^2\frac{\beta}{2})^{m-k}(\sin^2\frac{\beta}{2})^k-(\cos^2\frac{\beta}{2})^m\non\\
&=&1-(\cos^2\frac{\beta}{2})^m=1-|\bra\phi_1|\phi_2\ket|^{2m}.
\ee

The asymptotic limit above for $n\to\infty$ can be achieved in a different approach. As the number of copies in data system is infinite, the unknown state in it can be exactly reconstructed via quantum state tomography~\cite{Nielsen,Paris}. Denote the reconstructed state by $|\phi_0\ket$, and then $|\phi_0\ket$ is the state $|\phi_i\ket(i=1\ {\rm or}\ 2)$ with probability $\eta_i$. While the state in the data system is known, the discrimination between the total states comes to determining whether the state $|\phi_0\ket$ is equal to the state in program system $A$ or $C$. With $m$ copies of states in both program systems, we have $m$ pairs of states $|\phi_0\ket|\phi_1\ket$ and  $m$ pairs of  $|\phi_0\ket|\phi_2\ket$, and the task can now be completed by the state comparisons~\cite{PLA307.189} of the two states in the $2m$ pairs. More precisely, for each pair $|\phi_0\ket|\phi_i\ket (i=1\ {\rm or}\ 2)$, we project the state $|\phi_i\ket$ onto $|\phi_0^\bot\ket$, the state orthogonal to $|\phi_0\ket$ in the space spanned by $|\phi_1\ket$ and $|\phi_2\ket$, and the measurement result associated with the projection operator $E_0^\bot=|\phi_0^\bot\ket\bra\phi_0^\bot|$ gives the right answer of comparison. Thus, the failure probability of the comparison for each pair is $1-\bra\phi_i|E_0^\bot|\phi_i\ket=|\bra\phi_0|\phi_i\ket|^2$, and the success probability for at least one pair is $1-\prod_{i}|\bra\phi_0|\phi_i\ket|^{2m}=1-|\bra\phi_1|\phi_2\ket|^{2m}$, holding for both $|\phi_0\ket=|\phi_1\ket$ and $|\phi_0\ket=|\phi_2\ket$. Since the discrimination task succeeds unless the state comparisons of the $2m$ pairs of states all fail, the success probability via state comparisons is $P'_m=1-|\bra\phi_1|\phi_2\ket|^{2m}$,
which is explicitly the same as that in Eq.~(\ref{ninfty}).

(iii) For the averaged input states, the programmable unambiguous discriminator reduces to the usual unambiguous discrimination between known states as the number of copies in program systems goes to infinity~\cite{PRA73.012328,PRA82.042312}. The situation for the pure input sates has not been involved and now we can address this problem here. Since $n_A=n_C$, we have
\be
\label{minfty}
P=\sum_{k=1}^m\big(1-\eta_1q_{k,1}^{\rm opt}-\eta_2q_{k,2}^{\rm opt}\big) \sum_{l=-\frac{n_C}{2}}^{\frac{n_C}{2}-k}\bigg|D^{(\frac{n_C}{2})}_{l\frac{n_C}{2}}C^{\frac{N}{2}-k,\frac{n_1}{2}+l}_{\frac{n_1}{2}\frac{n_1}{2},\frac{n_C}{2}l}\bigg|^2,\non\\
\ee
where the relationship Eq.~(\ref{CG}) in the lemma has been used. With large $m$, using the normal approximation for the binomial distribution, one can obtain
$\big|D^{(\frac{n_C}{2})}_{l\frac{n_C}{2}}\big|^2\approx\frac{1}{\sqrt{2\pi}\sigma}\exp\big[-\frac{(m/2+l-\mu)^2}{2\sigma^2}\big]$,
which is a normal distribution with the expectation $\mu=m\cos^2(\beta/2)$ and the variance $\sigma^2=m\cos^2(\beta/2)\sin^2(\beta/2)$. Therefore, as $m\to\infty$, $
\big|D^{(\frac{n_C}{2})}_{l\frac{n_C}{2}}\big|^2\approx\frac{1}{m}\delta(\frac{l}{m}+\frac{1}{2}-\cos^2\frac{\beta}{2})$,
and then $l\approx l_0\equiv(\cos^2\frac{\beta}{2}-1/2)m$ is required for the nonzero terms in Eq.~(\ref{minfty}),
\be
P\approx\sum_{k=1}^m\big(1-\eta_1q_{k,1}^{\rm opt}-\eta_2q_{k,2}^{\rm opt}\big)\bigg|C^{\frac{N}{2}-k,\frac{n_1}{2}+l_0}_{\frac{n_1}{2}\frac{n_1}{2},\frac{n_C}{2}l_0}\bigg|^2.\non
\ee

To obtain the asymptotic limit of the CG coefficient $C^{\frac{N}{2}-k,\frac{n_1}{2}+l_0}_{\frac{n_1}{2}\frac{n_1}{2},\frac{n_C}{2}l_0}$ for large $m$, we first investigate the two expressions $\hat{\bm J}^2/m^2|(\frac{n_1}{2}\frac{n_C}{2})\frac{N}{2}-k,\frac{n_1}{2}+l_0\ket$ and $\hat{\bm J}^2/m^2|\frac{n_1}{2}\frac{n_1}{2},\frac{n_C}{2}l_0\ket$, where $\hat{\bm J}=\hat{\bm J}_{AB}+\hat{\bm J}_C$ is the total angular momentum for the whole system. Notice that $|(\frac{n_1}{2}\frac{n_C}{2})\frac{N}{2}-k,\frac{n_1}{2}+l_0\ket$ is the eigenvector of $\hat{\bm J}^2$ with the eigenvalue $(\frac{N}{2}-k)(\frac{N}{2}-k+1)$, we easily have (set $\hbar=1$)
\be
&&\hat{\bm J}^2/m^2\bigg|(\frac{n_1}{2}\frac{n_C}{2})\frac{N}{2}-k,\frac{n_1}{2}+l_0\bigg\ket\non\\
&\approx&(1-\frac{k}{m})^2\bigg|(\frac{n_1}{2}\frac{n_C}{2})\frac{N}{2}-k,\frac{n_1}{2}+l_0\bigg\ket.\non
\ee
Define $\hat{j}^{\pm}=\hat{j}^x\pm i\hat{j}^y$ for angular momentum operators, and then $\hat{\bm J}^2=\hat{\bm J}_{AB}^2+\hat{\bm J}_C^2+2\hat{J}_{AB}^z\hat{J}_C^z+\hat{J}_{AB}^+\hat{J}_C^-+\hat{J}_{AB}^-\hat{J}_C^+$. So, with some algebra, we can have
\be
\hat{\bm J}^2/m^2\bigg|\frac{n_1}{2}\frac{n_1}{2},\frac{n_C}{2}l_0\bigg\ket\approx\cos^2\frac{\beta}{2}\bigg|\frac{n_1}{2}\frac{n_1}{2},\frac{n_C}{2}l_0\bigg\ket.\non
\ee
In the derivation above, the angular momentum relationships $\hat{j}^{\pm}|JM\ket=\sqrt{J(J+1)-M(M\pm 1)}|JM\pm1\ket$ have been applied. Now, we know that the angular momentum vectors $|(\frac{n_1}{2}\frac{n_C}{2})\frac{N}{2}-k,\frac{n_1}{2}+l_0\ket$ and $|\frac{n_1}{2}\frac{n_1}{2},\frac{n_C}{2}l_0\ket$ are both the eigenvectors of the same Hermitian operator $\hat{\bm J}^2/m^2$, and therefore the CG coefficient $C^{\frac{N}{2}-k,\frac{n_1}{2}+l_0}_{\frac{n_1}{2}\frac{n_1}{2},\frac{n_C}{2}l_0}$ vanishes unless the two corresponding eigenvalues are equal, say $(1-k/m)^2\approx\cos^2(\beta/2)$ for large $m$. Further notice that $\sum_{k=0}^{n_C}\bigg|C^{\frac{N}{2}-k,\frac{n_1}{2}+l_0}_{\frac{n_1}{2}\frac{n_1}{2},\frac{n_C}{2}l_0}\bigg|^2=1$,
and we can conclude that $\bigg|C^{\frac{N}{2}-k,\frac{n_1}{2}+l_0}_{\frac{n_1}{2}\frac{n_1}{2},\frac{n_C}{2}l_0}\bigg|^2\approx \frac{1}{m}\delta(\frac{k}{m}-2\sin^2\frac{\beta}{4})$
as $m\to\infty$. Using the results above, Eq.~(\ref{minfty}) reduces to
\be
P\approx\big(1-\eta_1q_{k,1}^{\rm opt}-\eta_2q_{k,2}^{\rm opt}\big)\big|_{k/m=2\sin^2(\beta/4)}.\non
\ee
From Eq.~(\ref{parameter}), and together with $O_k\approx(1-k/m)^n$ for large $m$~\cite{QIC12.1017}, finally one has
\be
\label{minfinity}
P_n\equiv\lim\limits_{m \to \infty}{P}=\left\{\begin{array}{ll} \eta_2(1-\cos^{2n}\frac{\beta}{2}) & \textrm{for } \eta_1<e\\
1-2\sqrt{\eta_1\eta_2}\cos^n\frac{\beta}{2} & \textrm{for } e\leqslant\eta_1\leqslant f\\
\eta_1(1-\cos^{2n}\frac{\beta}{2}) & \textrm{for } \eta_1> f\end{array}\right.,\non\\
\ee
with the boundaries $e=\cos^{2n}\frac{\beta}{2}/(1+\cos^{2n}\frac{\beta}{2})$ and $f=1/(1+\cos^{2n}\frac{\beta}{2})$. 

The asymptotic limit for $m\to\infty$ in Eq.~(\ref{minfinity}) is exactly the same as that for the usual unambiguous discrimination between two known sates $|\phi_1\ket^{\otimes n}$ and $|\phi_2\ket^{\otimes n}$ with inner product $|{}^{\otimes n}\bra\phi_1|\phi_2\ket^{\otimes n}|=\cos^n(\beta/2)$. Actually, this asymptotic limit can indeed be approached by the usual unambiguous discrimination strategy here. As the numbers of copies in the program systems are both infinite, we can also reconstruct the unknown states in them via quantum state tomography, and the problem now comes to determining the states in the data system is $|\phi_1\ket^{\otimes n}$ or $|\phi_2\ket^{\otimes n}$.

(iv) Though receiving little attention before, the optimal measurement operators are important in the universal unambiguous discrimination. The expressions of the optimal measurement operators in Eq.~(\ref{operators}) reveal the symmetry properties in the universal unambiguous discrimination, which are rather useful in the derivation of our results in this paper. Before our work, the PSPs are achieved only for few simple examples since the general case is much more complicated without the explicit expressions of the optimal measurement operators. More significantly, the expressions of the optimal measurement operators theoretically represent the measurements in the experiments and with these operators, one can design the optimal universal unambiguous discriminators in the laboratory.

\textit{Conclusions and summaries.---} We derive the analytic expressions of PSPs for the universal programmable unambiguous discriminators using the explicit expressions of the optimal measurement operators given in Ref.~\cite{QIC12.1017}. Since pure states are actually input, PSPs are the success probabilities  when the device works, and we show that the optimal programmable unambiguous discriminator is equivalent to the usual unambiguous discrimination between the pure input states in the data register as the numbers of copies in program registers go to infinity, and equivalent to a series of state comparisons as the number of copies in data register goes to infinity. Similar conclusions hold for the averaged input states, and moreover, our result is more general, allowing for the arbitrary {\it a priori} probabilities $\eta_1$ and $\eta_2$, arbitrary dimension $d$ and arbitrary number of copies in the registers. It is the PSPs, instead of the ASPs, that are directly observed in laboratory, and hence the results in this paper are useful and helpful to the experimental realization of the universal unambiguous discriminators in the future. We expect that our results could come up with further theoretical or experimental consequences.

\textit{Acknowlegement.---} The author is grateful to the referee for very helpful comments.

\textit{Appendix: Proof of the lemma.---}
Firstly, notice that $|\phi_1\ket$ and $|\phi_2\ket$ are two states in $\mathcal{H}$, so there always exists a certain $d$-dimensional unitary transformation  $u\in{\rm U}(d)$ such that
\be
\label{qubitstates}
|\phi'_1\ket&=&u|\phi_1\ket=|0\ket,\non\\
|\phi'_2\ket&=&u|\phi_2\ket=e^{-\im\frac{\alpha+\gamma}{2}}\cos\frac{\beta}{2}|0\ket+e^{\im\frac{\alpha-\gamma}{2}}\sin\frac{\beta}{2}|1\ket,
\ee 
where $|0\ket$ and $|1\ket$ are two orthogonal bases of $\mathcal{H}$, and the unknown parameters $\alpha$, $\beta$ and $\gamma$ are Euler angles of the rotation group with $\cos\frac{\beta}{2}=|\bra\phi_1|\phi_2\ket|$,  $\beta\in[0,\pi]$ and $\alpha,\gamma\in[0,2\pi)$. Without any knowledge about $|\phi_1\ket$ and $|\phi_2\ket$, it is impossible to determine the exact expression of $u$, but it does exist, and this is enough. Now, we can define a new state $
|\Phi'_1\ket=u^{\otimes N}|\Phi_1\ket$.

Secondly, recall that $\bigg|\begin{array}{c} [\lambda]\\  \omega\end{array}\bigg\ket_1$$(\omega=1,2,...,d^{[\lambda]})$ are the irreducible bases for the representation labeled by $[\lambda]$, and then one can have
\be
u^{\otimes N}\openone^{[\lambda]}(u^{\otimes})^\dag=\sum_{\omega=1}^{d^{[\lambda]}}\bigg|\begin{array}{c} [\lambda]\\  \omega\end{array}\bigg\ket\bigg\bra\begin{array}{c} [\lambda]\\  \omega\end{array}\bigg|,
\ee
where the bases $\bigg|\begin{array}{c} [\lambda]\\  \omega\end{array}\bigg\ket(\omega=1,2,...,d^{[\lambda]})$ are the irreducible basis of both the permutation group $S_N$ and the unitary group U($d$), and can be transformed to $\bigg|\begin{array}{c} [\lambda]\\  \omega\end{array}\bigg\ket_1$ by the subduction coefficients (SDCs) of the permutation group $S_N$~\cite{Chenbook}. We then arrive at
\be
\label{identity}
\bra\Phi_1|\openone^{[\lambda]}|\Phi_1\ket=\sum_{\omega=1}^{d^{[\lambda]}}\bigg\bra\Phi'_1\bigg|\begin{array}{c} [\lambda]\\  \omega\end{array}\bigg\ket\bigg\bra\begin{array}{c} [\lambda]\\  \omega\end{array}\bigg|\Phi'_1\bigg\ket.
\ee

Thirdly, since $|\phi'_1\ket$ and $|\phi'_2\ket$ are in the space spanned by $|0\ket$ and $|1\ket$, they can be regarded as states of spin-$1/2$ system with $|0\ket$ and $|1\ket$ the spin-up and spin-down bases. Thus, for the state $|\Phi'_1\ket$, the system consists of $A$ and $B$ has the angular momentum $j_{AB}=n_1/2$, while the system $C$ has $j_C=n_C/2$.
Based on these, $|\Phi'_1\ket$ can be expressed as a linear combination of angular momentum basis
\be
\label{CG2}
|\Phi'_1\ket=\sum_{l=-\frac{n_C}{2}}^{\frac{n_C}{2}}D^{(\frac{n_C}{2})}_{l\frac{n_C}{2}}(\alpha,\beta,\gamma)|\frac{n_1}{2}\frac{n_1}{2}\ket|\frac{n_C}{2}l\ket,
\ee
with $\alpha$, $\beta$ and $\gamma$ the unknown parameters in Eq.~(\ref{qubitstates}). The bases $\bigg|\begin{array}{c} [\lambda]\\  \omega\end{array}\bigg\ket$ which contribute to Eq.~(\ref{identity}) should be the eigenstates like $|\frac{n_1}{2}\frac{n_C}{2}(\frac{N}{2}-k)M\ket$ with the total angular momentum $J=N/2-k$ and its $z$ component $M$, where $k=0,1,\cdots,n_C$ and $M=k-\frac{N}{2},k-\frac{N}{2}+1,\cdots,\frac{N}{2}-k$ for each $k$, and substituting this and Eq.~(\ref{CG2}) into Eq.~(\ref{identity}), one can have Eq.~(\ref{CG}).\qed


\begin{thebibliography}{99}
\bibitem{explain} $\mathcal{H}^{[n]}$ is the totally symmetric subspace of the $n$-partite Hilbert space $\mathcal{H}^{\otimes n}$, and $[n]$ is a Young diagram.
\bibitem{explain2} The symbols used for CG coefficients in this paper are $C^{jm}_{j_1m_1,j_2m_2}=\bra j_1m_1,j_2m_2|(j_1j_2)jm\ket$.
\end{thebibliography}
\end{document}